\documentclass[showpacs,preprintnumbers,amsmath,amssymb,12pt,floatfix]{revtex4}

\usepackage[usenames, dvipsnames]{color}
\usepackage{graphicx}
\usepackage{amssymb}
\usepackage[normalem]{ulem}
\renewcommand\sout{\bgroup\color[rgb]{1,0,0} \ULdepth=-.5ex \ULset}

\begin{document}
\title{Odd-even shape staggering and kink structure of charge radii of Hg isotopes by the deformed relativistic Hartree–Bogoliubov theory in continuum}
\author{Myeong-Hwan Mun} 
\affiliation{Department of Physics and Origin of Matter and Evolution of Galaxies (OMEG) Institute, Soongsil University, Seoul 06978, Korea}

\author{Seonghyun Kim}
\affiliation{Department of Physics and Origin of Matter and Evolution of Galaxies (OMEG) Institute, Soongsil University, Seoul 06978, Korea}

\author{W. Y. So} 
\affiliation{Department of Radiological Science, Kangwon National University at Dogye, Samcheok 25945, Korea}

\author{Myung-Ki Cheoun}
\email{cheoun@ssu.ac.kr (Corresponding Author)}
\affiliation{Department of Physics and Origin of Matter and Evolution of Galaxies (OMEG) Institute, Soongsil University, Seoul 06978, Korea}

\author{Soonchul Choi}
\affiliation{Center for Exotic Nuclear Studies, Institute for Basic Science, Daejeon 34126, Korea}

\author{Eunja Ha}
\address{Department of Physics and Research Institute for Natural Science, Hanyang University, Seoul, 04763, Korea}

\date{\today}

\begin{abstract}
We examined the shape staggering of relative charge radii in $^{180 - 186}$Hg isotopes, which was first measured in 1977 and recently confirmed using advanced spectroscopy techniques. To understand the nuclear structure underlying this phenomenon, we employed the deformed relativistic Hartree–Bogoliubov theory in continuum (DRHBc). Our analysis revealed that the shape staggering can be attributed to nuclear shape transition in the Hg isotopes. Specifically, we demonstrated that prolate shapes of $^{181,183,185}$Hg lead to an increase in the charge radii compared to oblate shapes of $^{180,182,184,186}$Hg isotopes.
We explained the nuclear shape staggering in terms of the evolution of occupation probability (OP) of $\nu 1 i_{13/2}$, $\nu 1 h_{9/2}$, $\pi 1 h_{9/2}$, and $\pi 3 s_{1/2}$ states. Additionally, we clarified the kink structure of the charge radii in the Hg isotopes near $N = 126$ magic shell does not come from the change of the OP of $\pi 1 h_{9/2}$ state, but mainly by the increase of the OPs of $\nu 1 i_{11/2}$ and $\nu 2 g_{9/2}$ states. 
\end{abstract}

\maketitle

\section{Introduction}

Experimental investigations of nuclear charge radii have revealed lots of intriguing regular patterns, abrupt changes, and non-linear trends along their isotopic chains in the nuclear chart. Charge radii are fundamental quantities that describe atomic nuclei and generally scale with either masses or charges. The development of rare isotope production science \cite{Kofo1951,Decr1991,Tani1985}, along with the significant advancements in detector technologies, particularly Multi-Reflection Time-of-Flight, Collinear Resonance Ionization Spectroscopy techniques (CRIS), and Resonance Ionization Laser Ion Source (RILIS) complex
\cite{Fedo2017} enabled high-precision measurements of nuclear masses and charge radii of nuclei far from the $\beta$-stability line. These measurements have been conducted on a range of isotopes, including K \cite{Rose2015,Krei2014,Pota1,Pota2}, Ca \cite{Wien2013,Garc2016,Ca}, Pd \cite{Geld2022}, Sn \cite{Gorg2019}, Hg \cite{Good2021,Mars2018,Kuehl1977,Ulm1986}, Pb \cite{Anse1986,Good2021}, and Bi \cite{Barz2021}.

Further, the odd-even shape staggering of charge radii is a unique property of odd-even nuclear isotopes, but the staggering was not significant in most nuclei except of the Hg isotopes \cite{Good2021,Mars2018,Kuehl1977,Ulm1986}. In order to describe the shape staggering of the odd Hg isotopes, Monte Carlo Shell Model (MCSM) \cite{Otsuka2020} and various density functional theories (DFTs) \cite{Sels2019} calculations have been performed with the largest particle model space.  The combination of the strong monopole interactions between $\pi 1 h_{9/2}$ state above $Z = 82$ magic shell and $\nu 1 i_{13/2}$ state in the mid-shell between $N = 82$ and $N = 126$, and the quadrupole interactions of nucleons near Fermi surfaces leading to the deformation turned out to cause the shape staggering phenomena. But it was argued that the results by DFT approach are very unlikely to reproduce the radius staggering \cite{Sels2019}.

Another recent calculation utilizing the Skyrme-Hartree-Fock model with pairing interaction by the BCS approximation reconfirmed that the observed staggering in $^{182 - 188}$Hg isotopes may arise from the prolate deformation of $^{181,183,185}$Hg and the oblate deformation of relevant nuclei \cite{More2022}. However, the explicit treatment of the Pauli blocking in odd nuclei was not performed in this calculation.    

Shape coexistence is another intriguing property in nuclear structure that has been extensively investigated in both theoretical nuclear models and experimental studies \cite{Pove2016,Gade2016,Heyde2011,Nach2004,Andr2000,Ojala2022,Paul2009}. In heavy nuclei, such as $^{184,186}$Pb isotopes, two-quasiparticle and four-quasiparticle configurations by protons can also lead to shape coexistence \cite{Waut1994, Duguet2003, Kim2022}. The energy spectrum of $^{186}$Pb indeed exhibits such shape coexistence in conjunction with quadrupole deformation ($\beta_2$) \cite{Andr2000}, providing a qualitative understanding of the experimental rotational band structures observed in $^{186}$Pb \cite{Duguet2003}. We note that recent data \cite{Ojala2022} confirmed the shape coexistence by the electric monopole transitions among the excited and ground $0^+$ states in the isomer related to the shape coexistence. Not only the Pb isotopes, the shape coexistence in $^{181,183,185}$Hg has been established in the last century \cite{Varm1997,Bind1993,Zhang2022}.

Another notable phenomenon known as the kink structure, which represents a rapid increase of charge radii above a magic shell, has been observed across shell closures \cite{Krei2014}. For instance, a kink structure has been observed in K and Ca isotopes near the magic shell $N = 20$ \cite{Pota1,Ca}. 
Concurrently, recent data for Hg isotopes from the in-source RILIS at CERN-ISOLDE have revealed a distinct kink at $N = 126$, along with remarkable shape staggering observed around $N = 100 - 106$ \cite{Good2021,Mars2018}. 
These characteristics present a complementary approach for investigating the evolution of shell structure as well as nuclear shape transition in nuclear isotopes. The observed kink structure in the Hg isotopes is believed to be a consequence of the swelling of the neutron core mainly by the $\nu 2 g_{9/2}$ state\cite{Good2021,Rong2022,Hori2022,Dong2023}. 

In this work, we provide evidences that the observed shape staggering in the Hg isotopes is a manifestation of shape coexistence within the framework of the DRHBc. This theory has proved to be a reliable systematic framework for describing fundamental nuclear properties across the entire nuclear chart, including regions near the neutron- and proton-drip lines \cite{Kai2020,ADND2022,Cong2022}. Firstly, we present the shape coexistence of Hg isotopes  $(N = 99 - 109)$ by analysing the total binding energy (TBE) curves. These curves reveal the presence of multiple minima corresponding to different shapes within the isotopic chain. Secondly, we argue that the staggering observed in the Hg isotopes is closely linked to the shape coexistence by examining the occupation probabilities of the particles, $\nu 1 i_{13/2}$ and $\pi 1 h_{9/2}$ states, around neutron Fermi energy. The analysis demonstrates a strong correlation between the staggering and the presence of shape coexistence. Furthermore, we establish that the observed kink structure is intimately connected to the swelling of the neutron core at $N = 126$ followed by the pulling of the protons by the symmetry energy.


\section{Formalism}

To address aforementioned issues comprehensively, the development of a refined relativistic nuclear model becomes imperative. This model should incorporate deformation, pairing correlations, and the continuum within a microscopic framework capable of explaining the entire nuclear mass range. In this regard, the DRHBc theory has been developed. This theory encompasses both meson-exchange density functionals and point-coupling density functionals, and has been specifically designed to describe deformed halo nuclei \cite{Zhou2010,Zhou2012}. More recently, it has been successfully extended to address even-odd nuclei through the use of an automatic blocking method \cite{Cong2022}.
The DRHBc theory has exhibited remarkable predictive power in the description of nuclear masses \cite{ADND2022,Pan2021,Kaiyuan2021}, and auspiciously applied to nuclei in close proximity to the drip lines \cite{Pan2019,Sun2020,Sun2018,Yang2021,Sun2021,Sun2021-2,Sun2021-3}.

In this work, we focus on the shape coexistence and its subsequent effect, namely, shape staggering and kink structure, in the Hg isotopes. We begin by considering the following Hartree-Bogoliubov equation:

\begin{equation} \label{eq:hfbeq}
\left( \begin{array}{cc} h_D - \lambda &
\Delta  \\
 - \Delta^{*} & - h_D^{*} + \lambda
  \end{array}\right)
\left( \begin{array}{c}
U_{k} \\ V_{k}  \end{array}\right)
 =
 E_{k}
\left( \begin{array}{c} U_{k} \\
V_{k} \end{array}\right),
\end{equation}
where $h_D, \lambda, E_k, U_k$, and $V_k$ are Dirac Hamiltonian, Fermi energy, quasiparticle energy and wave functions, respectively.
In coordinate space, the Dirac Hamiltonian is
\begin{equation}\label{eq:ham}
h_D ({\bf r}) = {\bf \alpha} \cdot {\bf p} + V ({\bf r}) + \beta [M + S({\bf r})]~,
\end{equation}
where $M$ is the nucleon mass, and $S({\bf r})$ and $V ({\bf r})$ are the scalar
and vector potentials, respectively, which can be derived from an effective Lagrangian in Ref. \cite{Cong2022}.
The resulting potentials are given in terms of local scalar, vector and isoscalar densities ($\rho_S, \rho_V$ and $\rho_3$)
\begin{eqnarray}\label{eq:pot}
S({\bf r}) &=& \alpha_S \rho_S + \beta_S \rho_S^2 + \gamma_S \rho_S^3 + \delta_S \Delta \rho_S, \\ \nonumber
V({\bf r}) &=& \alpha_V \rho_V + \gamma_V \rho_V^3 + \delta_V \Delta \rho_V + e A^0
+ \alpha_{TV} \tau_3 \rho_3 + \delta_{TV} \tau_3 \Delta \rho_3 ~,
\end{eqnarray}
where the subscripts S, V , and T stand for
scalar, vector, and isovector, respectively. The isovector-scalar
channel including the terms $\alpha_{TS}$ and $\delta_{TS}$ in Eq. (\ref{eq:pot}) is neglected
since including the isovector-scalar interaction does not improve the description of nuclear ground-state properties \cite{Burv2002}.

The paring potential $\Delta$ is determined by the pairing tensor $\kappa ({\bf r}, {\bf r}^{'})$ and is given by:
\begin{equation} \label{eq:Delta}
\Delta({\bf r}, {\bf r}^{'}) = V ({\bf r}, {\bf r}^{'}) \kappa ({\bf r}, {\bf r}^{'})~,~
V({\bf r}, {\bf r}^{'}) = {V_0 \over 2} ( 1 - P_{\sigma}) \delta ( {\bf r} - {\bf r}^{'}) ( 1 -  { \rho (\bf r) \over \rho_{sat}} )~.
\end{equation}

To account for even-odd nuclei, we consider the blocking effect of unpaired nucleon(s) by treating the ground state as one-quasiparticle state as follows \cite{Cong2022}
\begin{equation}
| \Phi_1 > = \beta_{k_b}^{\dagger} | \Phi > = \beta_{k_b}^{\dagger} \Pi_{k} \beta_{k (\neq k_b)} | 0 >,
\end{equation}   
where $\beta_{k_b}^{\dagger}$ is the quasiparticle creation operator for the properly
blocked state $b$. The blocked orbital
$k_b$ causing the time-odd mean field \cite{Poto2010} breaks the time reversal symmetry, produces the currents, and the single-particle state with +$m$ and its conjugate
state with -$m$ are no longer degenerate. For simplicity, the
equal filling approximation (EFA) is usually adopted \cite{Li2012,Martin2008},
by which the currents vanish and the two configurations of a
particle in the $\pm m$ space are averaged in a statistical manner. Then we obtain in each
step of the iteration the fields with the time reversal symmetry.
Correspondingly, the density matrix $\rho$ and pairing tensor $\kappa$ in
Eq. (\ref{eq:Delta}) are replaced by \cite{Li2012,Suga1966,Ring1970,Sasa2003}
\begin{eqnarray}
\rho^{'} &=& \rho	 +{1 \over 2} (U_{k_b} {U^{\ast T}}_{k_b} - V^{\ast}_{k_b} {V^{T}}_{k_b}) \\ \nonumber
\kappa^{'} &=& \kappa	 +{1 \over 2} (U_{k_b} {V^{\ast T}}_{k_b} - V^{\ast}_{k_b} {U^{T}}_{k_b}) ~.
\end{eqnarray}
For the description of the ground states as well as the excited states, the particle vibration coupling (PVC) due  to the dynamical property of nuclei could play an important role because it affects the single particle state energies given by a mean field \cite{Litvi2011,Niu2015}.   
In odd nuclei, the PVC effect could be significant by the coupling of odd-nucleon to the remained core \cite{Hama1969,Good2021}. For instance, the PVC lowers the energy of $\nu 2 g_{9/2}$ state below  that of the $\nu 1 i_{11/2}$ state \cite{Good2021}. But the exact treatment of the PVC in the present DRHBc model is beyond the scope of this work, we do not consider the PVC effect.
 
Finally, the nucleon energy part in total energy of odd nucleus is given as 
\begin{equation}
E_{nucleon} = 2  \sum_{{k>0 (k \neq k_b) ,m>0}} (\lambda - E_k) v_k^2 + (\lambda + E_{k_b}) u_{k_b}^2 - (\lambda - E_{k_b}) v_{k_b}^2 - 2 E_{pair}~,
\end{equation}
where the second and third term stand for the blocked nucleon energy with the projection of the blocked state, respectively, $m >$ 0 and $m <$ 0. In our numerical calculation, we employed automatic blocking, which refers to the blocking of the lowest quasiparticle orbitals instead of the orbital-fixed blocking. Remarkably, the results obtained using the automatic blocking for nuclei such as $^{23}$Mg and $^{22}$Al were found to be nearly identical to those by the orbital-fixed blocking \cite{Cong2022}. This demonstrates the consistency and reliability of the automatic blocking method in the present numerical calculations.

\begin{figure}
\centering
\includegraphics[width=0.90\linewidth]{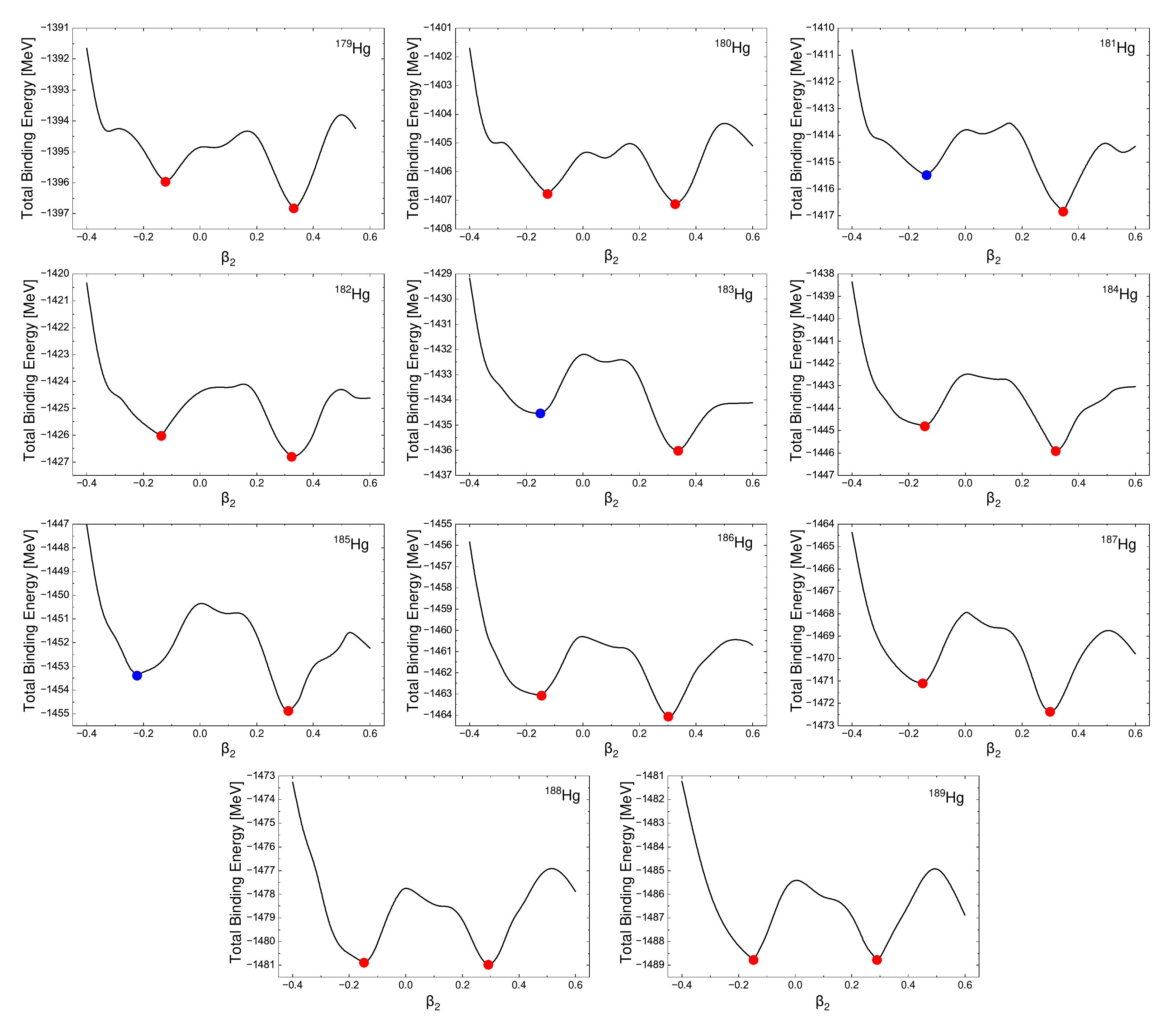}
\caption{(Color online) TBE curves in terms of $\beta_2$ for $^{179-189}$Hg isotopes. All of the isotopes demonstrate a possibility of the shape coexistence coming from about 1 MeV energy difference between strong prolate and weak oblate deformation. The blue filled circles in $^{181,183,185}$Hg disclose a local minimum in the oblate deformation region, which are about 1 MeV higher than those in prolate deformation.}
\label{fig1}
\end{figure}

In this work, we use a pairing strength of $V_0$ = -- 325.0 MeV fm${^3}$ with a pairing window of 100 MeV, and adopt a saturation density of $\rho_{sat}$ = 0.152 fm$^{-3}$. The energy cutoff $E_{cut}^+ =$ 300 MeV, the angular momentum cutoff $J_{max} = (23/2) \hbar $, and the Legendre expansion truncation as ${\lambda}_{max}$ = 8 \cite{Kai2020,Kaiyuan2021} are taken for the Dirac Woods-Saxon basis, which were adopted for the DRHBc mass table \cite{ADND2022,Cong2022}.

\section{Results}

First, we acknowledge the preference of oblate deformation over prolate deformation in Hg isotopes with $A \ge 190$, as previously established \cite{ADND2022}. However, when examining the TBE curves of $^{179-189}$Hg isotopes in Fig.\ref{fig1}, we find that both deformations exhibit nearly identical TBEs. This suggests the possibility of shape coexistence in this mass region. In fact, $^{181,183,185}$Hg isotopes are known to have shape coexistence by observing the fragment mass-$\gamma$ and $\gamma-\gamma$ coincidence from five prolate deformed rotational bands \cite{Bind1993,Varm1997}.

However, it is important to note that, for $^{181,183,185}$Hg, the TBE differences in the oblate deformation (blue circles) are a bit larger than those of other odd and even isotopes (red circles). This observation has significant implications for the subsequent discussion on the shape staggering in the Hg isotopes \cite{Mars2018}.

\begin{figure}
\centering
\includegraphics[width=0.99\linewidth]{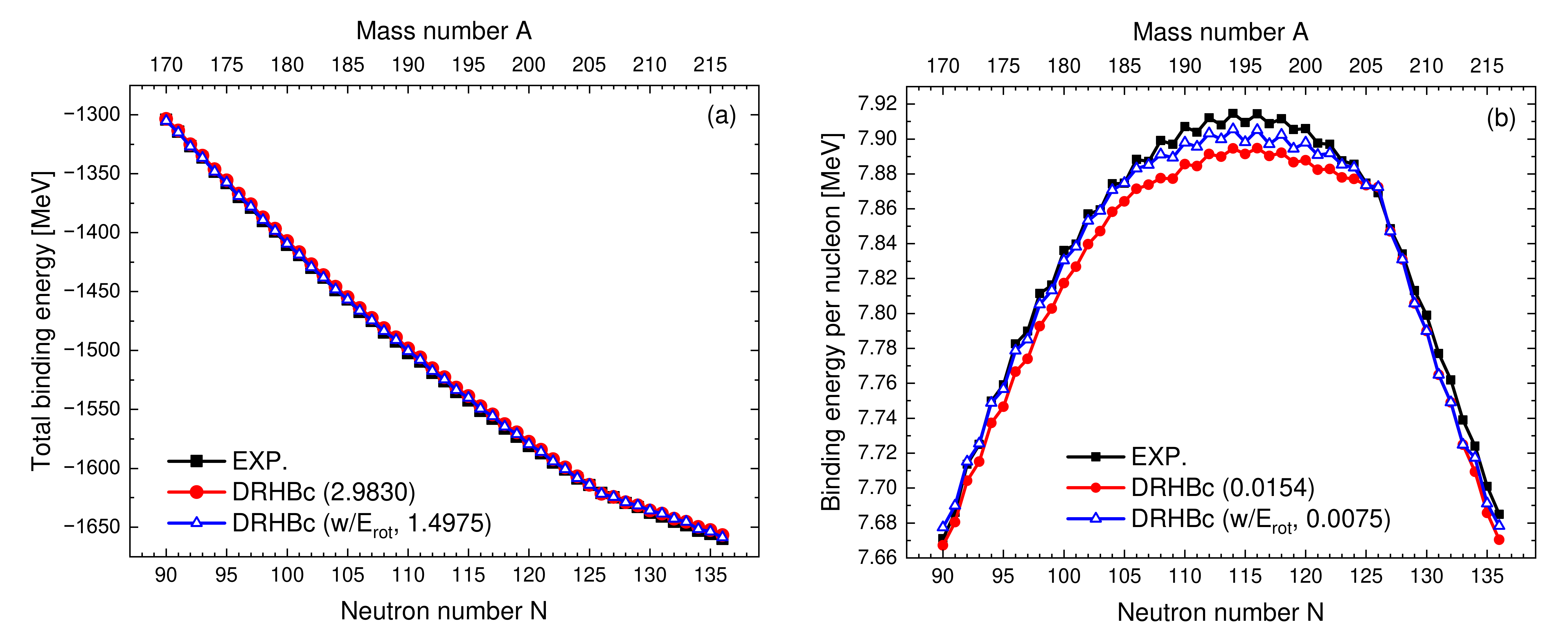}
\caption{(Color online) TBEs (a) and BE/A (b) of even and odd Hg isotopes ($N = 90 - 136$) from  DRHBc calculations with (blue) and without (red) the rotational energies. The numbers in parenthesis stand for average deviation to the data \cite{AME2020}.}
\label{fig2}
\end{figure}

In Fig.\ref{fig2}, we present the TBEs (a) and binding energy per nucleon (BE/A) (b) obtained from the DRHBc theory \cite{ADND2022}, along with the corresponding experimental data \cite{AME2020}. Both the TBEs and BE/A show good agreement to the experimental data, with uncertainties less than 1.5 MeV and 0.02 MeV, respectively, in the TBE and BE/A. This agreement provides a reliable basis for investigating the shape staggering and kink structure in the Hg isotopes.

\begin{figure}
\includegraphics[width=0.45\linewidth]{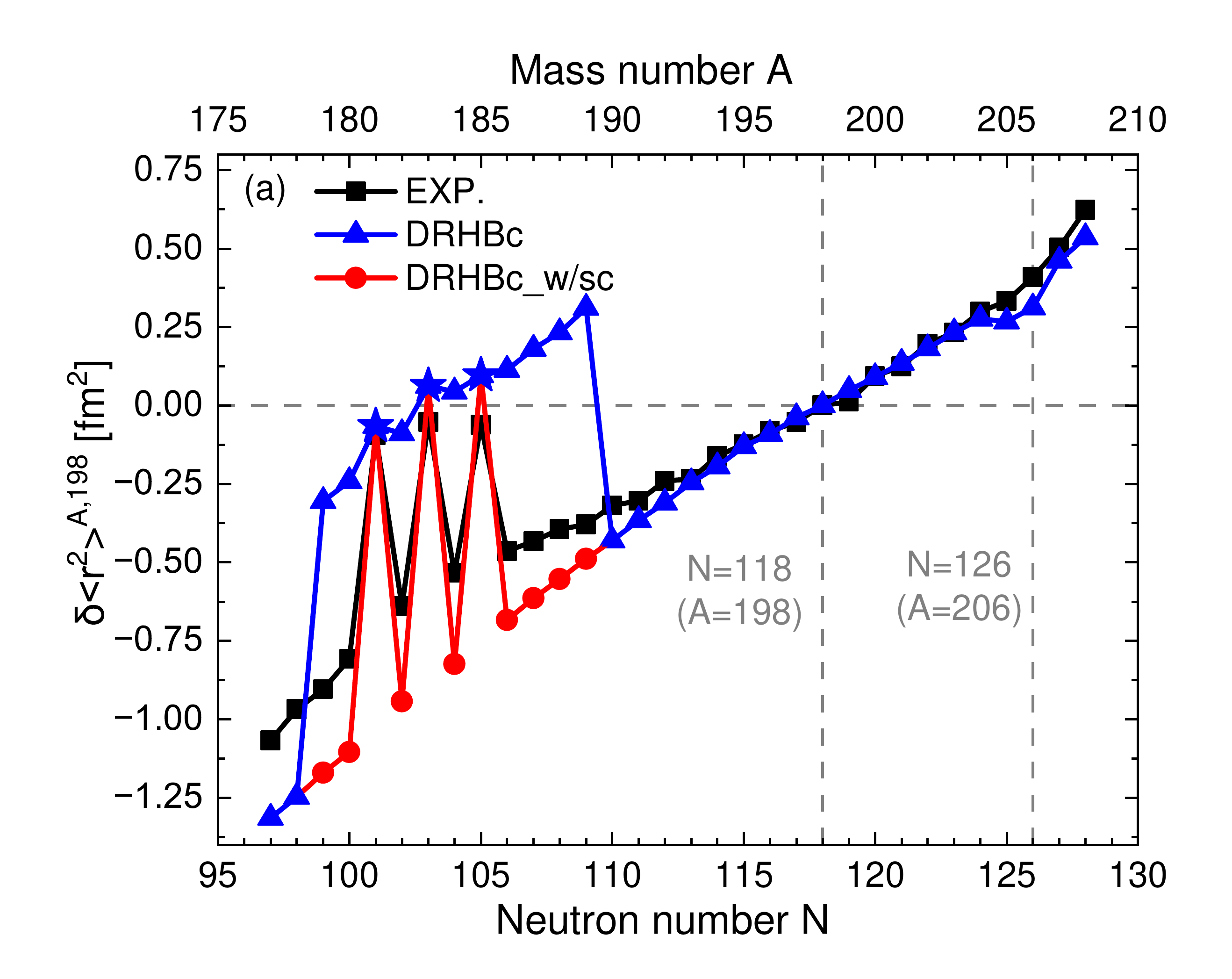}
\includegraphics[width=0.52\linewidth]{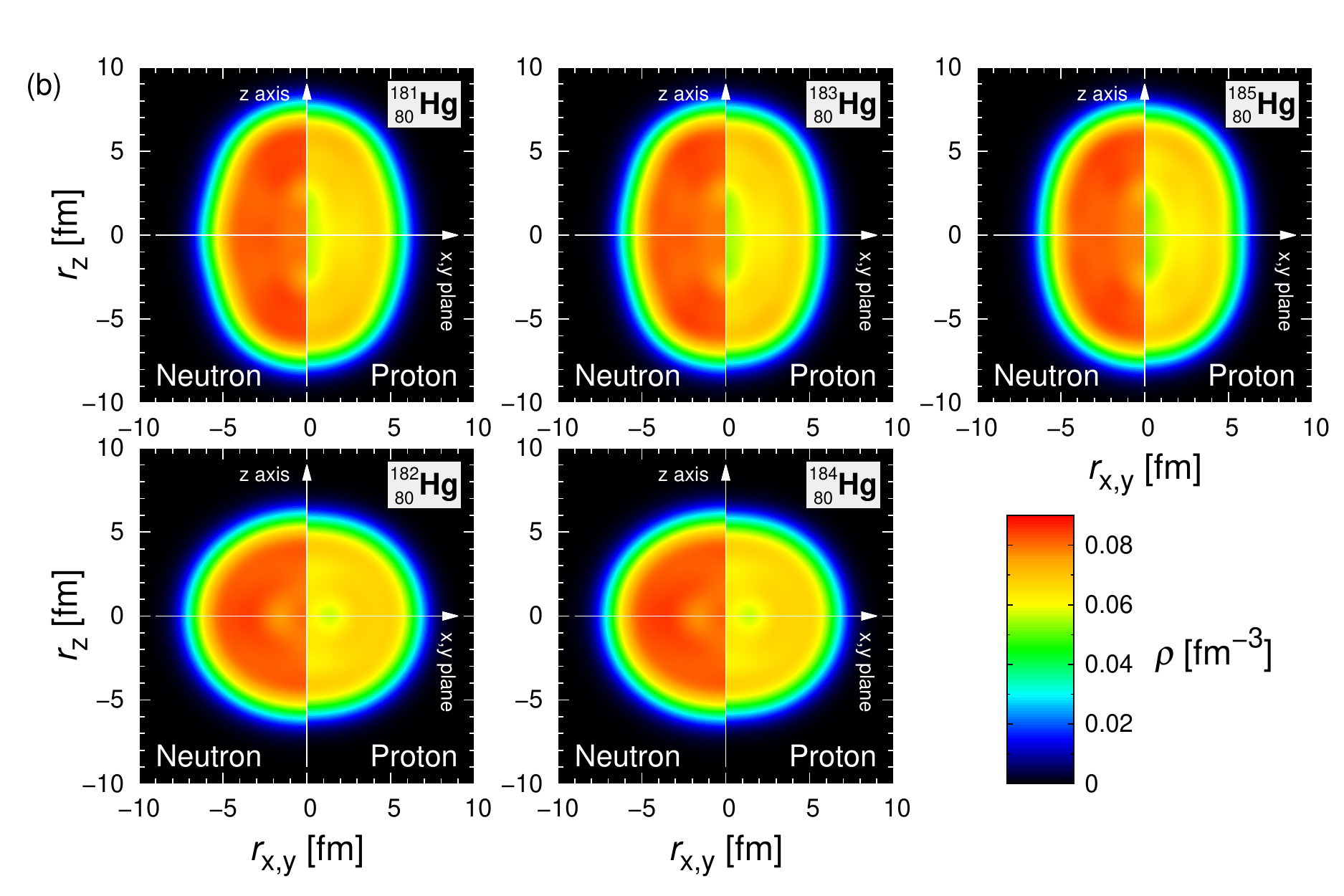}
\caption{(Color online) Relative changes of mean square charge radii $\delta {< r^2 >}^{A,198}$ of the Hg isotopes. Black boxes stand for the data \cite{Good2021}. Results of $N \geq 110$ ($N \leq 109$) (blue color) are obtained by oblate deformation (prolate deformation) shapes, while red colors are calculated by oblate deformation considering the shape coexistence taken from Fig.\ref{fig1}. For $^{181,183,185}$Hg, blue stars denote the results by the prolate deformation obtained in Fig.\ref{fig1}. The contours illustrate the neutron (left half) and proton (right half) prolate shapes of $^{181,183,185}$Hg (upper) and oblate shapes of $^{182,184}$Hg (lower) by their density distributions.} 
\label{fig3}
\end{figure}

Moving forward, in Fig.\ref{fig3}(a), we discuss the shape staggering and kink structure in the Hg isotopes. We present the experimental data and theoretical results obtained by calculating relative changes in mean square charge radii, $\delta {<r^2>}^{A, A'}~=~<r^2 (A)> - <r^2 (A')>~=~r^2_{ch} (A) - {r^2}_{ch} (A')$, with respect to the $A' =~^{198}$Hg isotope. The results by oblate deformation for $N \geq 110$ (blue triangles) show good agreement with the experimental data (black squares), capturing the kink structure observed in the charge radii \cite{ADND2022}. However, for the $N = 99 - 109$ region, the prolate deformation results (blue triangles and stars) overestimate the charge radii data, except for the odd $^{181,183,185}$Hg isotopes. Considering the presence of shape coexistence, the results obtained using oblate deformation (red circles) as suggested in Fig.\ref{fig1} reasonably account for the relative charge radii data within $\delta {<r^2>} \leq$ 0.25 fm$^2$. Nevertheless, the charge radii data for the $^{181,183,185}$Hg isotopes (black squares) cannot be explained by the oblate deformation.

\begin{figure}
\includegraphics[width=0.48\linewidth]{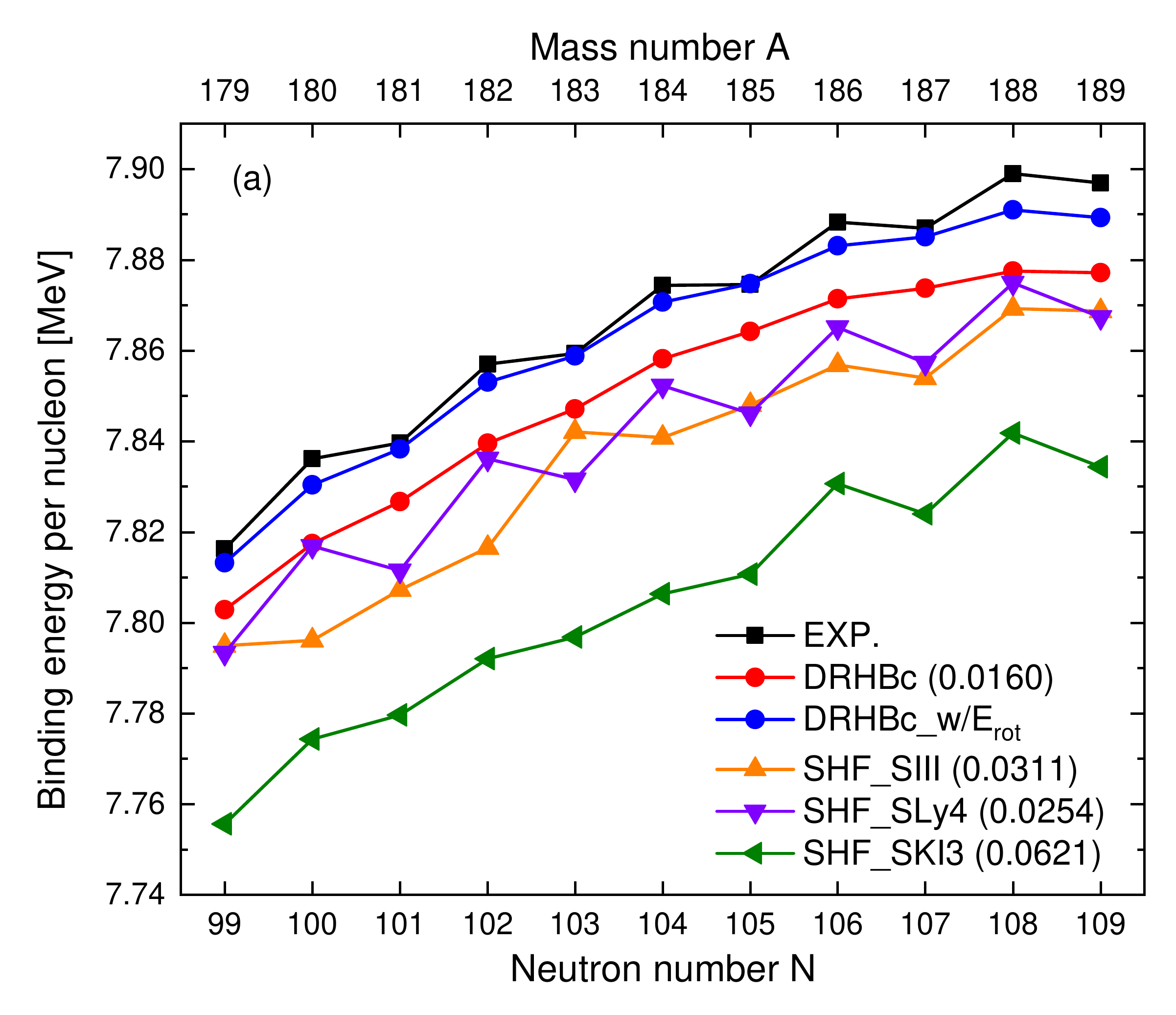}
\includegraphics[width=0.48\linewidth]{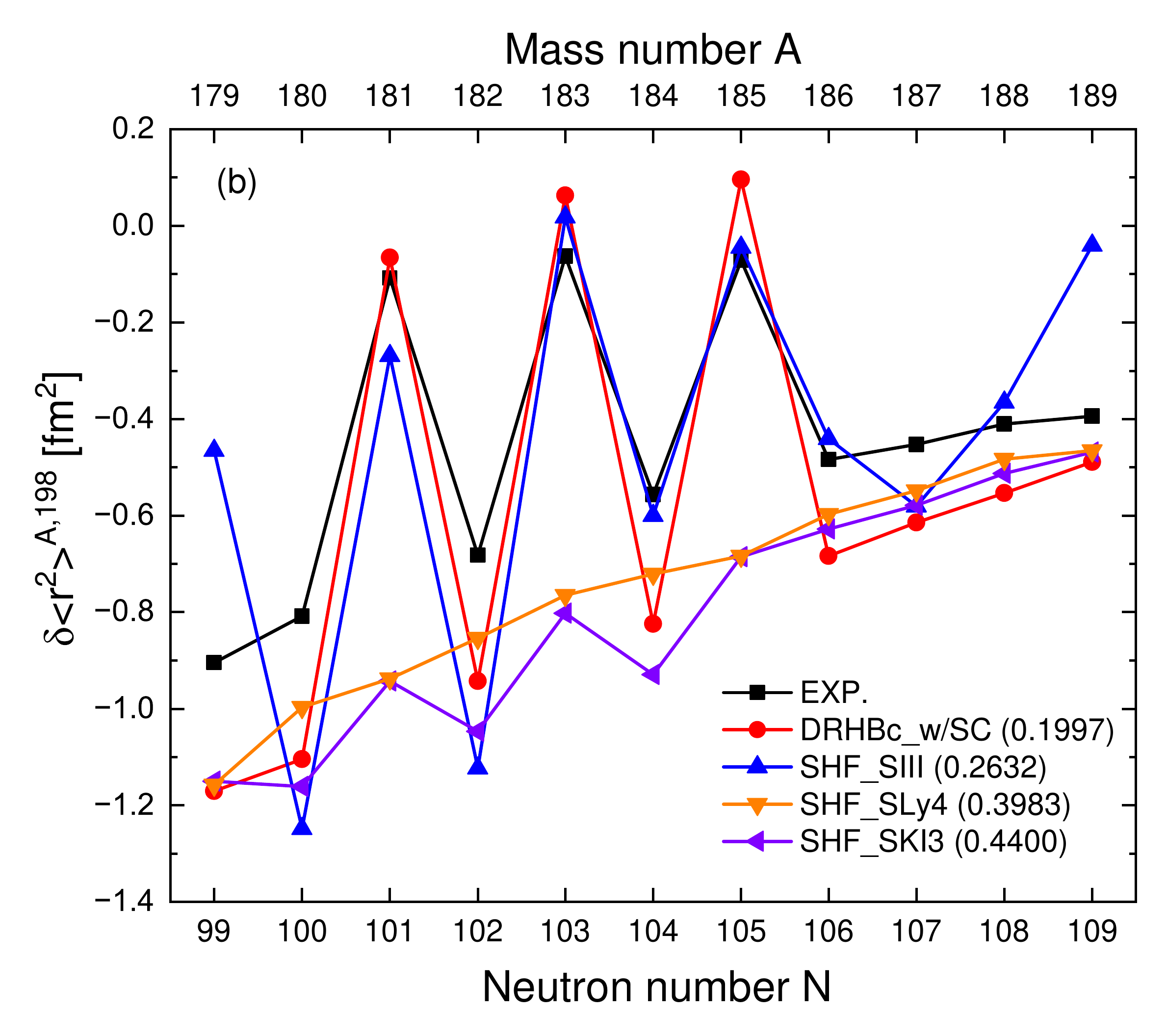}
\caption{(Color online) Comparison of DRHBc and Skyrme models for the BE/A (a) and shape staggering (b) corresponding to Fig.\ref{fig2} (b) and Fig.\ref{fig3} (a). The experimental data are taken from Refs. \cite{AME2020,Good2021}.} 
\label{fig4}
\end{figure}

\begin{table}[h]
\caption{Deformations of Hg isotopes used in Fig. 3 and 4 by DRHBc and various Skyrme DFTs.}
\begin{tabular}{c c c c c c}
\hline
\hline
                & DRHBc       & SIII \cite{SIII}     & SKI3 \cite{SkI3}     & SLy4 \cite{SLy4}    & Exp. ($\beta_2$ by $Q_2$) \\
\hline
 $^{179}$Hg    & $-0.1210$    & $ 0.2751$    & $ 0.1139$    & $ 0.1039$     & $-$ \\
 $^{180}$Hg    & $-0.1250$    & $ 0.0000$    & $ 0.0000$    & $-0.1372$     & $0.1380$ \\
 $^{181}$Hg    & $ 0.3458$    & $ 0.2854$    & $-0.1321$    & $-0.1380$     & $-$ \\
 $^{182}$Hg    & $-0.1360$    & $ 0.0000$    & $ 0.0000$    & $-0.1437$     & $0.1470$ \\
 $^{183}$Hg    & $ 0.3367$    & $ 0.3033$    & $-0.1441$    & $-0.1498$     & $-$ \\
 $^{184}$Hg    & $-0.1430$    & $-0.1732$    & $ 0.0000$    & $-0.1469$     & $0.1560$ \\
 $^{185}$Hg    & $ 0.3118$    & $ 0.2714$    & $-0.1432$    & $-0.1420$     & $-$ \\
 $^{186}$Hg    & $-0.1460$    & $-0.1790$    & $-0.1462$    & $-0.1467$     & $0.1310$ \\
 $^{187}$Hg    & $-0.2400$    & $-0.1379$    & $-0.1439$    & $-0.1454$     & $-$ \\
 $^{188}$Hg    & $-0.1450$    & $-0.1698$    & $-0.1436$    & $-0.1441$     & $0.1450$ \\
 $^{189}$Hg    & $-0.1464$    & $ 0.2268$    & $-0.1400$    & $-0.1259$     & $-$ \\
 $^{198}$Hg    & $-0.1164$    & $-0.1088$    & $-0.1147$    & $-0.1085$     & $0.1064$ \\
\hline
\hline
\end{tabular}
\label{Deformations}
\end{table}

However, since the local minima in the TBE curves of $^{181,183,185}$Hg isotopes are located in prolate region, as emphasized in Fig.\ref{fig1}, we maintain prolate deformation. This choice leads to charge radii of $^{181,183,185}$Hg (blue stars) that align with the larger radii data \cite{Good2021}. We admit that there appear some discrepancies to the charge radii data, but they are maximally 0.05 fm \cite{ADND2022}. 
\begin{figure}
\centering
\includegraphics[width=1.0\linewidth]{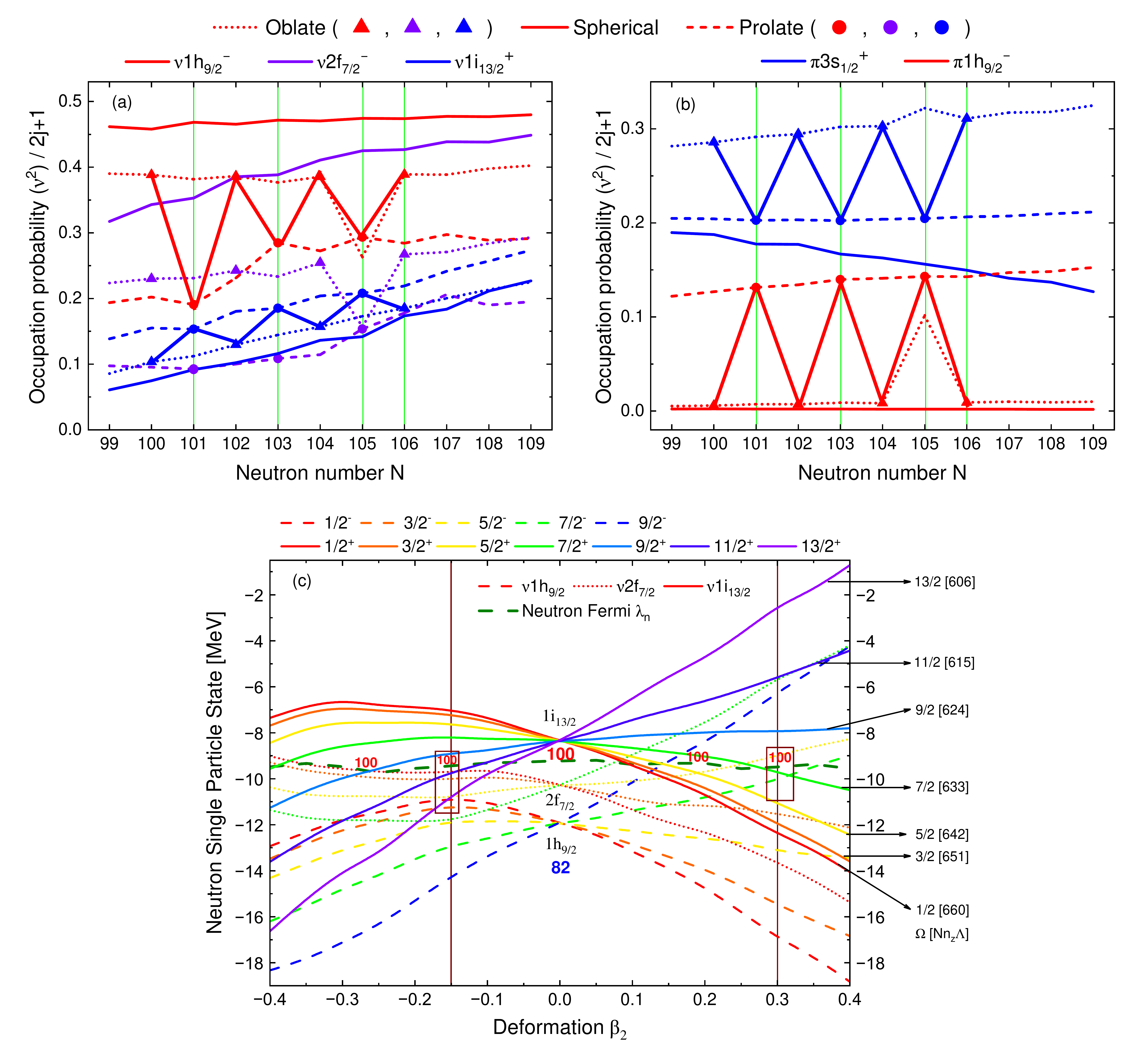}
\caption{(Color online) Evolution of occupation probabilities of the neutron (a) and proton (b) SPSs for the Hg isotopes. In prolate deformation region, the occupation probabilities of $\nu 1h_{9/2}$, $\nu 2 f_{7/2}$, and $\pi 3 s_{1/2}$ states decrease while those of ${\nu 1i_{13/2}}$ and $\pi 1 h_{9/2}$ states increase. This behaviour can be understood by the shell evolution for $\beta_2$ with Nilsson quantum numbers $\Omega (N,n_z,\Lambda)$ in the panel (c).}
\label{fig5}
\end{figure}

Indeed, the shape staggering in the Hg isotopes is a result of the shape transition from prolate deformation to oblate deformation due to the presence of shape coexistence in the Hg isotopes. In particular, the odd isotopes $^{181,183,185}$Hg maintain prolate deformation (blue stars) because of the local minima in the prolate region. The shape transition in the $^{181-185}$Hg isotopes is clearly demonstrated by the density distribution contours in Fig.\ref{fig3} (b), which exhibit a distinct change in nuclear shape from prolate to oblate. Additionally, it is noteworthy that the proton deformations closely resemble those of the neutrons in these isotopes.

\begin{figure}
\centering
\includegraphics[width=1.0\linewidth]{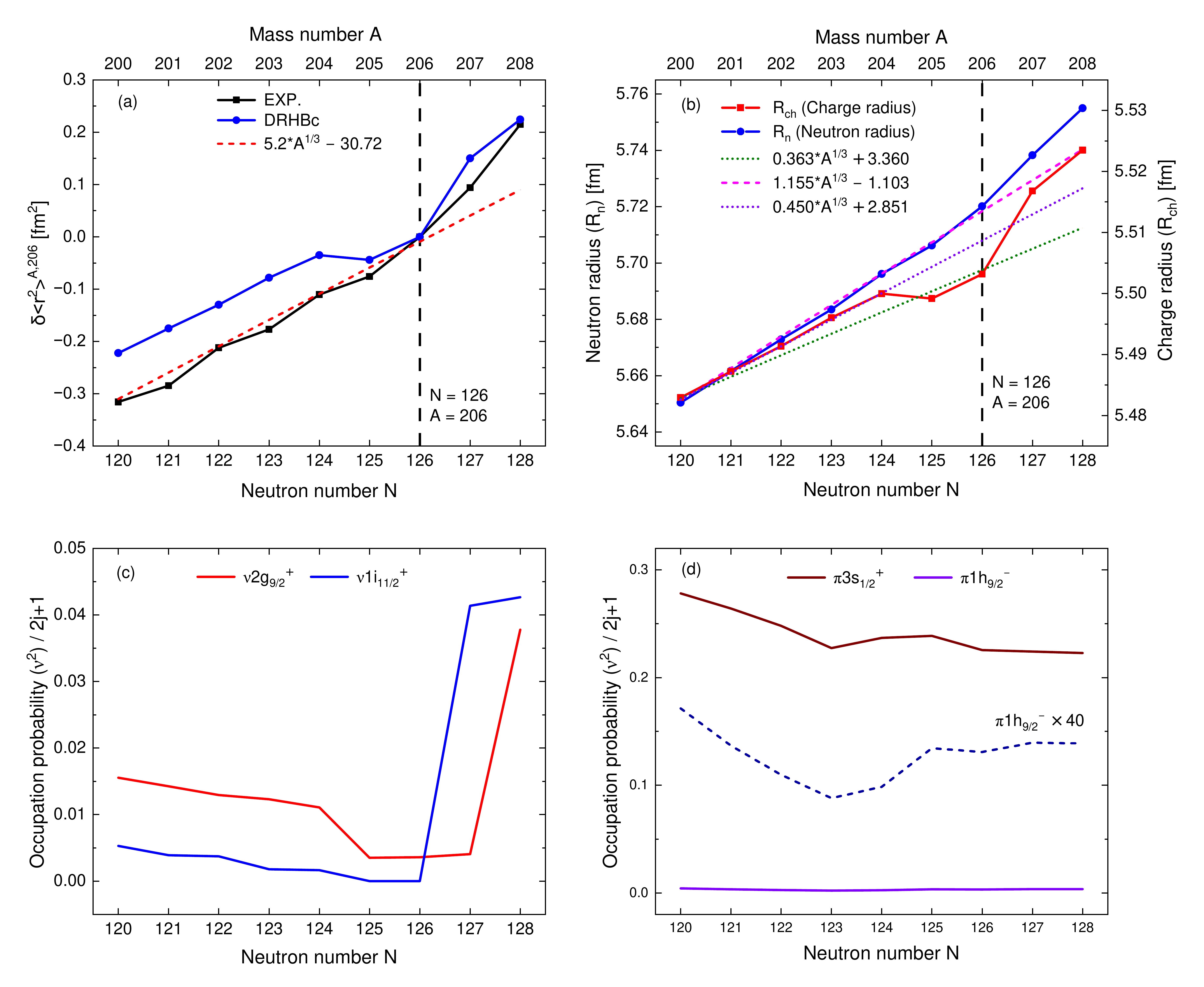}
\caption{(Color online) Kink structure around N = 126 shell by the relative radius difference (a) and $R_n$ with $R_{ch}$ (b) w.r.t. $^{206}$Hg. The dashed lines are just from the continuation of the data before N = 126 case to show the kink structure. Panel (c) shows the larger increase in the occupation probabilities of $\nu 1 i_{11/2}$ state than $\nu 2 g_{9/2}$ state. Occupation probabilities of proton states (d) are rarely changed.}
\label{fig6}
\end{figure}

The kink structure observed in the vicinity of $N = 126$ shell in Fig.\ref{fig3} (a) is another intriguing feature. Although the data points depend on the choice of the reference nucleus (see Fig.\ref{fig6} (a)), the results obtained from the DRHBc model calculations exhibit a clear kink structure that agrees with the trend of experimental data. In the subsequent discussion, we will delve into the physics underlying the shape staggering and kink structure, taking into consideration the shell evolution of the main single-particle-states (SPSs) near the Fermi energy.

In Fig.\ref{fig4}, we also demonstrate the shape staggering by the non-relativistic Skyrme DFT using some parameter sets \cite{SIII, SLy4, SkI3}. The results by the present DRHBc model is superior to the results by other Skyrme DFTs. For the shape staggering, one model (SIII \cite{SIII}) provides reasonable staggering pattern of odd Hg isotopes. In Table I, we present the deformation parameters obtained from each model. The results by SIII model shows the nuclear shape transition similar to the present results by DRHBc model. It means that the nuclear shape transition is the most important aspect for understanding the shape staggering of the Hg isotopes. Here we note that recent calculations by Fayans DFT interpret the shape staggering by strong prolate for odd isotopes and weakly prolate deformations \cite{Boro2023}.

In Fig.\ref{fig5} (a), we present the shell evolution of occupation probabilities of neutron SPSs for ${\nu 1 i_{13/2}}$, ${\nu 2 f_{7/2}}$, and ${\nu 1 h_{9/2}}$ states around the Fermi energy for $N = 100$ in the Hg isotopes. The triangles represent the oblate deformation for even isotopes, while the circles represent the prolate deformation for the odd Hg isotopes. Notably, we observe an abrupt change in the occupation probability of ${\nu 1 i_{13/2}}$ state in the prolate deformed $^{181,183,185}$Hg isotopes, which confirms previous findings obtained from MCSM calculations \cite{Mars2018}. Conversely, we observe decrease in the occupation probability of ${\nu 1 h_{9/2}}$ and ${\nu 2 f_{7/2}}$ states.

Similarly, in Fig.\ref{fig5} (b), we observe similar patterns in the proton occupation probabilities for $\pi 3s_{1/2}$ (decrease) and $\pi 1 h_{9/2}$ (increase) states around $Z = 80$. This implies that $\pi 1 h_{9/2}$ state plays a crucial role in the abrupt increase of the charge radii of the odd Hg isotopes. These abrupt increases  can be attributed to the quadrupole component of the nucleon-nucleon interaction as well as the monopole interaction between ${\nu 1 i_{13/2}}$ and ${\pi 1 h_{9/2}}$ states. Furthermore, the pairing interactions in even-even nuclei contribute to the lower binding energies and the reduction of charge radii in even isotopes. 

The observed seesaw-like change in occupation probabilities can be attributed to the evolution of neutron SPSs. Figure\ref{fig5} (c) illustrates the evolution of relevant states with $\beta_2$. In the larger prolate deformation region, $\nu 1 i_{13/2}$ substates with lower $\Omega$ values below the Fermi surface (green dashed line) are more occupied compared to those in the oblate deformation region, while the number of $\nu 1 h_{9/2}$ substates decreases.

Moving on to the discussion of the kink structure observed in the vicinity of the $N = 126$ magic shell \cite{Good2021}, this phenomenon is a prominent feature observed in magic shell nuclei such as K, Ca, and Pb. In our $\delta < r^2 >$ and $R_n$ with $R_{ch}$ results for the Hg isotopes in Fig.\ref{fig6} (a) and (b), we observe a kink structure above $N = 126$ magic shell, characterized by the swelling of neutron radii, although there still remained some discrepancy about $\delta <r^2> \sim$ 0.1. This swelling effect is responsible for the observed kink structure. As previously mentioned \cite{Hori2022}, pairing correlations lead to the expansion of the neutron core, resulting in the increase in the occupation probabilities of $\nu 2 g_{9/2}$ and $\nu 1 i_{11/2}$ states beyond the N = 126 magic shell (Fig.\ref{fig6} (c)).

On the other hand, there is no corresponding increase in the occupation probabilities of $\pi 3s_{1/2}$ and $\pi 1 h_{9/2}$ states in the proton SPS spectra (Fig.\ref{fig6} (d)). This supports the notion that the kink structure is primarily driven by the swelling of neutrons, particularly, of $\nu 1 i_{11/2}$ state, not by the increase of the proton states such as $\pi 3 s_{1/2}$ and $\pi 1 h_{9/2}$ states. The monopole interactions by the spin-triplet force between protons and neutrons play a role in driving the protons to the surface region \cite{Godd2013}.

\section{Summary and Conclusion}
 
Our findings highlight the significance of shape coexistence in understanding the shape staggering observed in the Hg isotopes. It is clear that taking into account oblate deformation for the Hg isotopes, considering the shape coexistence for the $^{179-189}$Hg isotopes, is crucial for explaining the shape staggering phenomenon. The three prolate-shaped odd nuclei, $^{181,183,185}$Hg, are typical shape coexistence nuclei confirmed by the rotational band structure data and exhibit larger charge radii, indicating that the shape staggering is closely related to the shape coexistence in the Hg isotopes, or vice versa. By using the DRHBc framework, TBEs of the $^{181,183,185}$Hg isotopes are found in the prolate region, which clearly leads to the shape staggering. The seesaw-like occupation probabilities of corresponding single-particle-states in the $^{180-186}$Hg isotopes turn out to be essential for understanding the prominent shape staggering observed in the Hg isotopes.

As our calculation is based on the ground state properties, a proper understanding of nuclear deformation and shape coexistence would require considering quadrupole excitations beyond the mean-field approach, such as the collective Hamiltonian method or generator coordinate method. These methods can handle shape coexistence and provide new average deformation parameters and charge radii \cite{Yang2023}. A recent systematic calculation of E0 transitions among the $0^+$ ground states, incorporating the shape coexistence within the covariant DFT \cite{Yang2023}, could provide a more rigorous test of shape coexistence and the associated shape transitions as well as the small discrepancies in the present results. However, we note that this calculation was limited to even-even nuclei. This could be a direction for future work.

Furthermore, the DRHBc model employed in the present study successfully reproduces the kink structure observed around the N = 126 shell. This kink structure arises from the increase in the occupation probabilities of ${\nu i_{11/2}}$ and ${\nu 2 g_{9/2}}$ states beyond the magic shell, influenced by the multi-particle and multi-hole interactions. This swelling effect pushes the proton states to the outer nuclear surface without change of proton occupation probabilities by the symmetry energy residing on the symmetric core surface. It would be interesting to further investigate the shape staggering and the kink structures of charge radii and matter radii in K, Ca, Pb and Au isotopes in the vicinity of each magic shell using the DRHBc model in future studies.

\begin{acknowledgements}
Helpful discussions with members of the DRHBc Mass Table Collaboration are gratefully appreciated. In particular, we appreciate helpful comments and careful reading of the manuscript by Prof. A. N. Andreyev and Prof. Shuangquan Zhang.
M.H.M. and M.K.C. were supported by the National Research Foundation of Korea (NRF) grant funded by the Korea government (Grant Nos. NRF-2021R1A6A1A03043957 and NRF-2020R1A2C3006177). 
M.H.M. was also supported by the NRF (Grant Nos. NRF-2021R1F1A1060066 and NRF-2021R1F1A1046575). 
E.H. was supported by the NRF (Grant No. NRF-2018R1D1A1B05048026). 
W.S. was supported by the NRF (Grant No. NRF-2021R1F1A1046575). 
The work of SC is supported by the Institute for Basic Science (IBS-R031-D1). 
This work was supported by the National Supercomputing Center with supercomputing resources including technical support (KSC-2022-CRE-0333). 
\end{acknowledgements}

\end{document}